\documentclass[aps,pre,reprint,footinbib,superscriptaddress,floatfix,bm]{revtex4-1}
\def\arXiv{1}  
\if\arXiv1
\pdfoutput=1
\fi

\usepackage{amsmath,amsfonts,amssymb,bm}
\usepackage[dvipdfmx]{graphicx}
\usepackage[usenames]{color}
\usepackage[dvipsnames]{xcolor}
\usepackage{txfonts}
\usepackage{xspace}
\usepackage{hyperref}
\hypersetup{
  colorlinks = true,
  citecolor  = blue,
  linkcolor  = blue,
  urlcolor   = blue
}

\newcommand{\JA}{J_{\rm A}}
\newcommand{\absJA}{\lvert{J_{\rm A}}\rvert}
\newcommand{\JB}{J_{\rm B}}
\newcommand{\JAB}{J_{\rm AB}}
\newcommand{\etaeff}{\eta^{\rm eff}}
\newcommand{\heff}{H_{\rm eff}}

\begin{document}

\title{%
  Ordering phenomena in a heterostructure of frustrated and unfrustrated triangular-lattice Ising layers
}
\author{Milan \v{Z}ukovi\v{c}}
\email{milan.zukovic@upjs.sk}
\affiliation{Institute of Physics, Faculty of Science, P. J. \v{S}af\'arik University, Park Angelinum 9, 040 01 Ko\v{s}ice, Slovakia}
\author{Yusuke Tomita}
\email{ytomita@shibaura-it.ac.jp}
\affiliation{College of Engineering, Shibaura Institute of Technology, Saitama 337-8570, Japan}
\author{Y. Kamiya}
\affiliation{Condensed Matter Theory Laboratory, RIKEN, Wako, Saitama 351-0198, Japan}

\date{\today}

\begin{abstract}
  We study critical and magnetic properties of a bilayer Ising system consisting of two triangular planes A and B, with the antiferromagnetic (AF) coupling $J_{\rm A}$ and the ferromagnetic (FM) one $J_{\rm B}$ for the respective layers, which are coupled by the interlayer interaction $J_{\rm AB}$ by using Monte Carlo simulations. When $J_{\rm A}$ and $J_{\rm B}$ are of the same order, the unfrustrated FM plane orders first at a high temperature $T_{c1} \sim J_{\rm B}$. The spontaneous FM order then exerts influence on the other frustrated AF plane as an effective magnetic field, which subsequently induces a ferrimagnetic order in this plane at low temperatures below $T_{c2}$. When short-range order is developed in the AF plane while the influence of the FM plane is still small, there appears a preemptive Berezinskii-Kosterlitz-Thouless-like pseudocritical crossover regime just above the ferrimagnetic phase transition point, where the short-distance behavior up to a rather large length scale exponentially diverging in $\propto \JA / T$ is controlled by a line of Gaussian fixed points at $T = 0$. In the crossover region, a continuous variation in the effective critical exponent $4/9 \lesssim \eta^{\rm eff} \lesssim 1/2$ is observed. The phase diagram by changing the ratio $J_{\rm A}/J_{\rm B}$ is also investigated.
\end{abstract}

\maketitle

\section{Introduction}

Magnetism in thin films (i.e., bilayers and multilayers) is a rapidly developing research field due to their novel magnetic properties different from bulk materials as well as recent advances in their fabrication and characterization techniques at atomic scale~\cite{jong76,bala08,shi15}. This can lead to useful technological applications such as high-density magnetic recording and magnetic sensors~\cite{shim92}. One of the main theoretical interests lies in the possibility to study the crossover phenomena between two-dimensional (2D) and 3D systems~\cite{cape76,jong90}.

A number of previous studies focused on magnetic properties of simple Ising bilayers formed by two ferromagnetic (FM) layers coupled by an exchange interaction of varying strength~\cite{ferr91,hans93,hori97,lipo98,li01,kim01,ghae04,monr04,szal12,szal13}. Such bilayers have been shown to undergo phase transitions that belong to the 2D Ising universality class and their critical temperature is controlled by the shift exponent that depends on the interlayer to intralayer coupling ratio. In a recent numerical work on the thin-film Ising systems, composed of multiple layers, a systematic continuous deviation has been reported not only for the critical temperature but also for the critical exponents, the latter of which however could be more adequately regarded as effective exponents, relative to their values for the single layer system~\cite{phu09a}.

If the underlying lattice has frustrated in-plane interactions, the corresponding stacked system can have more nontrivial physics. It is well known that, in contrast to its ferromagnetic counterpart, a 2D triangular lattice Ising antiferromagnet (TLIA) shows no long-range order (LRO) phase down to zero temperature due to high geometrical frustration~\cite{wann50}. On the other hand, a 3D system obtained by stacking of individual TLIA planes on top of each other has been confirmed to display LRO at finite temperatures through the order-by-disorder mechanism~\cite{berker,blank,copper}.
Generally, frustrated spin systems can display remarkable and often unexpected properties (see, e.g., Ref.~\onlinecite{diep13} for a recent review). Peculiar critical behavior has been reported in the frustrated thin-film spin systems with Ising-like anisotropy~\cite{ngo07,phu09b}, showing crossover from the first- to second-order transition and two phase transitions related to disorderings of surface and interior layers, respectively. More recently, it was found that the interplay between the in-plane frustration and kink excitations fluctuating along the out-of-plane direction can induce ``stiffness from disorder'' phenomena in a layered system of a finite number of TLIA planes~\cite{lin14}. 

Motivated by these studies, in this paper we consider a bilayer system of classical Ising spins corresponding to a heterostructure of two triangular planes, the spins within which are coupled by antiferromagnetic (AF) interactions in one layer and FM interactions in the other. The two planes are coupled by the interlayer interaction, which we can assume either FM or AFM without loss of generality. As discussed above, the critical behavior of the individual planes is very different. While the FM one displays a phase transition in the Ising universality class to the FM LRO phase, the AF one shows no LRO down to zero temperature due to high geometrical frustration. A prior account of such a bilayer system has been provided in Ref.~\onlinecite{zuko16}, which pointed to the existence of the ferrimagnetic (FR) LRO phase also in the AF plane, induced by an effective field coming from the FM plane. In the present study we demonstrate that the competing ordering and disordering tendencies enforced by the respective unfrustrated and frustrated planes in the AF/FM bilayer result in a rather intricate critical and pseudocritical behaviors in the exchange interaction parameter space.

\section{Model and simulation details}
\subsection{Model}
The Hamiltonian of the bilayer Ising system [Fig.~\ref{fig:par_spc}(a)] is
\begin{align}
  \mathcal{H} = 
  -\JA\sum_{\langle i \in {\rm A},j \in {\rm A} \rangle}\sigma_{i}\sigma_{j}
  - \JB\sum_{\langle k \in {\rm B},l \in {\rm B} \rangle}\sigma_{k}\sigma_{l}
  - \JAB\sum_{\langle i \in {\rm A},k \in {\rm B} \rangle}\sigma_{i}\sigma_{k},
  \label{Hamiltonian}
\end{align}
where $\sigma_{i}=\pm 1$ is an Ising spin on the $i$th lattice site. The first (second) sum runs over nearest neighbors (NN) within the plane A (B), where $\JA < 0$ and $\JB > 0$, respectively, are the AF and the FM interactions in each plane. The third sum runs over NN between the planes A and B, coupled by the FM interaction $\JAB>0$. 
In this work, we take $\JB + \JAB$ as the unit of energy, unless otherwise specified.

\subsection{%
  \label{(sub)sec:parameterization}
  Parameter regime of our investigation
}

\begin{figure}
  \centering
  \if\arXiv0
  \includegraphics[width=\hsize]{FIG01_alt_w_lattice.eps}
  \else
  \includegraphics[width=\hsize,bb=0 0 629 219]{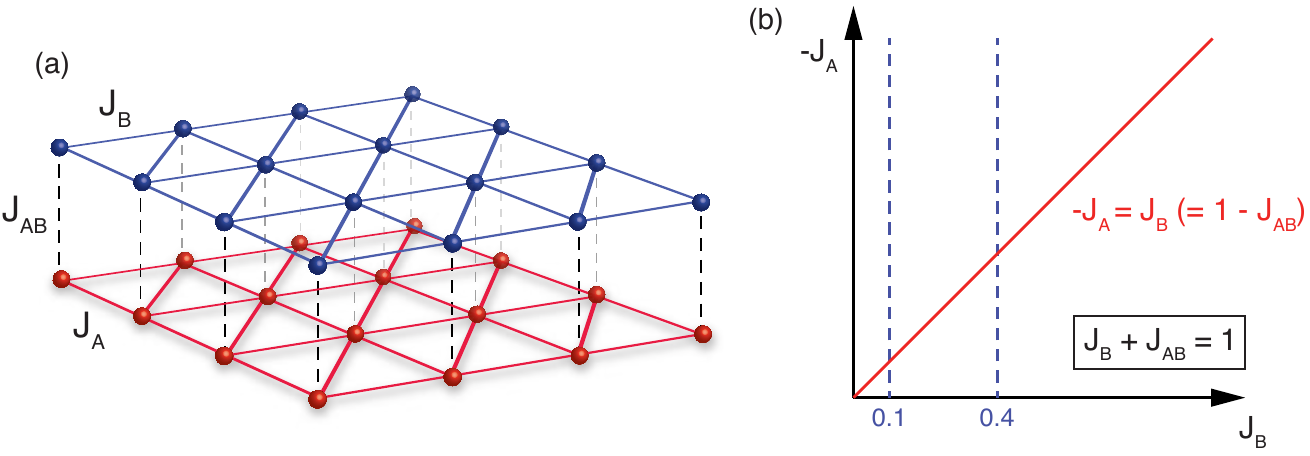}
  \fi
  \caption{%
    \label{fig:par_spc}
    \if\arXiv0
    (color online) 
    \fi
    (a) Structure of the bilayer triangular lattice.
    (b) The investigated regions of the parameter space. The solid red line and the dashed blue lines represent the case(s) with $-\JA = \JB$ and $-\JA \ne \JB$, respectively.
  }
\end{figure}

In the 2D parameter space $(\JA, \JB, \JAB)$ with $\JB + \JAB$ fixed, we select a couple of representative 1D cuts for our investigation presented in Sec.~\ref{sec:results}. First, we will consider the intraplane exchange interactions to have equal strengths and the intra- to interplane exchange interaction ratio will vary from zero to infinity, i.e., $-\JA=\JB=1-\JAB \equiv J$ with $J\in [0,1]$. This is a continuation of the previous investigation of the same model presented in Ref.~\onlinecite{zuko16}, where $J = 0.5$ was assumed (i.e., $-\JA=\JB=\JAB$). We will also consider more general cases where $\JA$ and $\JB$ have different amplitudes, taking $(\JB, \JAB) = (0.4, 0.6)$ and (0.1, 0.9) with varying $\JA$ as our examples. These 1D parameterizations are illustrated in Fig.~\ref{fig:par_spc}(b).

\subsection{Simulation details}
The model~\eqref{Hamiltonian} is studied by Monte Carlo (MC) simulations by using the standard Metropolis algorithm. We consider spin systems with the total number of sites $L\times L \times 2$, with $L=$ 24, 48, 72, 96, and 120, and apply periodic boundary conditions in the in-plane directions. To evaluate various thermodynamic quantities (see below), we typically consider $10^5$ Monte Carlo sweeps (MCS) for sampling after discarding $2 \times 10^4$ MCS for thermalization. In order to shorten the thermalization period at low temperatures, we start our simulation from a high temperature in the paramagnetic region with a random spin configuration and gradually decrease the temperature $T$ with a small step (typically $\Delta T=0.05$ or $0.02$, which is in units of $\JB + \JAB$ and the Boltzmann constant $k_{\rm B} \equiv 1$); the simulation at the next lower temperature starts from the final configuration obtained at the previous temperature.

\begin{figure}[t]
  \centering
  \if\arXiv0
  \includegraphics[width=0.65\hsize]{FIG02_tauint_OP1_J04.eps}
  \else
  \includegraphics[width=0.65\hsize,bb=0 0 393 334]{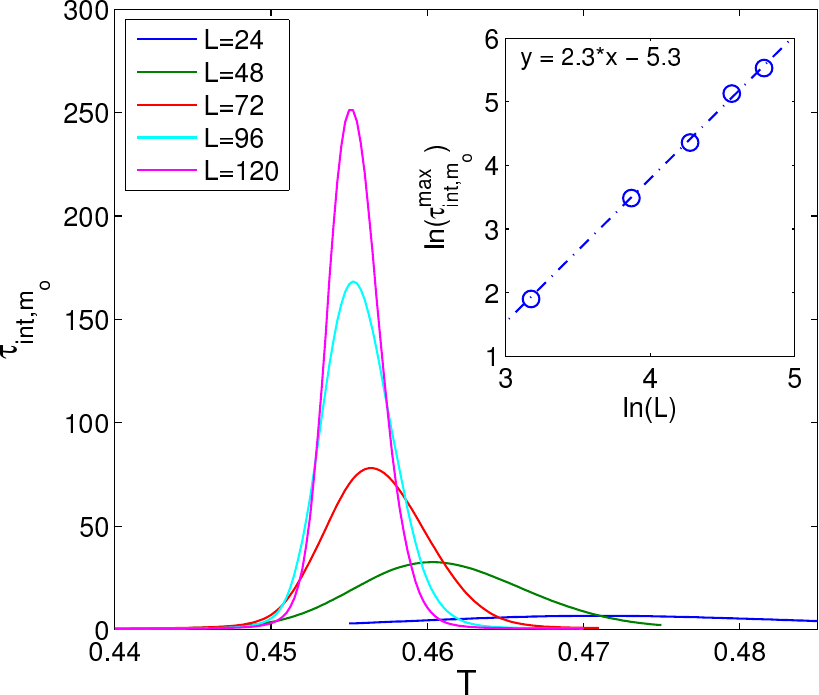}
  \fi
  \caption{%
    \label{fig:tauint_OP1_J04}
    \if\arXiv0
    (color online) 
    \fi 
    Integrated autocorrelation time $\tau_{{\rm int},m_o}$ near the low-temperature phase transition in the AF layer 
    for ($\JA$, $\JB$, $\JAB$) = (-0.4, 0.4, 0.6) with different lattice sizes $L$.
    The inset shows the $L$-dependence of the peak height showing a power law behavior with the estimated dynamic critical exponent $z \approx 2.3$.
  }
\end{figure}

In order to obtain the critical exponents, we perform a finite-size scaling (FSS) analysis, in which we elaborate our MC simulations with a larger number of MCS ($10^7$ MCS for sampling and $2 \times 10^6$ MCS for thermalization where the sampling is made every 10th MCS to reduce autocorrelation effects). We also apply the reweighting techniques~\cite{ferr88} to examine the temperature dependence in detail. We note that the autocorrelation is particularly enhanced near the low-temperature phase transition in the frustrated AF layer, and thus relatively long simulations are necessary to obtain a reliable output. As shown in Fig.~\ref{fig:tauint_OP1_J04}, the integrated autocorrelation time $\tau_{{\rm int},m_o}$ for the FR order parameter $m_o$ of the AF plane (see the definition below) follows $\tau_{{\rm int},m_o} \propto L^z$ with $z \approx 2.3$ and can be as large as the order of $10^2$ MCS for the largest lattice we studied. Our protocol for the FSS analysis ensures that the data quality is good enough for assessing the critical behavior at low temperatures.

\subsection{Observables}
We evaluate the following quantities, where $\langle\cdots\rangle$ denotes thermal averaging. The internal energy per site is 
\begin{equation}
  \label{ene}
  E = \frac{\langle {\mathcal H} \rangle}{2L^2},
\end{equation}
and
\begin{equation}
  \label{eq.c}
  C =\frac{\langle {\mathcal H}^{2} \rangle - \langle {\mathcal H} \rangle^{2}}{2L^2T^{2}}
\end{equation}
is the specific heat per site. From the $T$-dependence of the internal energy, we can derive the entropy density by using the thermodynamic integration method~\cite{kirk77} as
\begin{equation}
  \label{ent}
  S(T) = \ln 2
  + \frac{E(T)}{T}
  +  
  \int_{\infty}^{T} \frac{E(\tilde{T})}{\tilde{T}^2} d\tilde{T}.
\end{equation}
The magnetization per site in each plane is
\begin{equation}
  \label{mag}
  m_{\mathrm{A(B)}} 
  = \frac{\langle M_{\mathrm{A(B)}}\rangle}{L^2} 
  = \frac{1}{L^2} 
  \Big\langle\Bigl\lvert\!\sum_{i \in \mathrm{A(B)}}\sigma_{i}\Bigr\rvert\Big\rangle.
\end{equation}
We also define the three-sublattice FR order parameter within the AF plane A as
\begin{equation}
  \label{mag_o}
  m_{o} = \frac{\langle M_{o} \rangle}{L^2} 
  = \frac{1}{\sqrt{6}L^2}\biggl\langle\!\! \sqrt{\left(O^{(1)}\right)^2+\left(O^{(2)}\right)^2+\left(O^{(3)}\right)^2}\, \biggr\rangle,
\end{equation}
with $O^{(\mu)} = \sum_R \phi^{(\mu)}_R$ ($\mu=1,2,3$). Here, the summation runs over the enlarged unit cell $R$ comprising three spins in the plane A and
\begin{equation}
  \phi^{(1)}_R = \sigma^{}_{R,1} - \frac{1}{2} \left( \sigma^{}_{R,2} + \sigma^{}_{R,3} \right)\
  \label{eq:phi}
\end{equation}
is the FR local order parameter, where $ \sigma^{}_{R,\mu}$ ($\mu=1,2,3$) denote the $\mu$th sublattice spin of the enlarged unit cell at $R$ in the plane A ($\phi^{(2)}_R$ and $\phi^{(3)}_R$ are defined by cyclic permutation of the indices). We will refer to $\phi^{(1)}_R$ simply as $\phi^{}_R$ in what follows.
In addition, we define the Binder parameter associated with the FR order parameter:
\begin{align}
  U_{4,o}= 1-\frac{\langle m_{o}^{4}\rangle}{3\langle m_{o}^{2}\rangle^{2}}.
\end{align}

We also calculate several derivatives of $M_x$, $x=\mathrm{A}$, $\mathrm{B}$, and $o$. First, the susceptibility per site $\chi_{x}$, corresponding to the parameter $M_x$, is
\begin{equation}
  \label{eq.chi}
  \chi_{x} = \frac{\langle M_{x}^{2} \rangle - \langle M_{x} \rangle^{2}}{L^2T}, 
\end{equation}
and the derivative of $\langle m_x \rangle$ with respect to $1/T$ and the logarithmic derivatives of $\langle m_x \rangle$ and $\langle m_{x}^{2} \rangle$ with respect to the same parameter are
\begin{subequations}
\begin{align}
  \label{eq.D0}
  d^{}_{1/T}\,m_x &= -T^2 \frac{\partial}{\partial T}\langle m_x \rangle = \langle m_x {\mathcal H} \rangle- \langle m_x \rangle \langle {\mathcal H} \rangle,
  \\
  \label{eq.D1}
  d^{}_{1/T}\ln m_x &= -T^2 \frac{\partial}{\partial T}\ln\langle m_x \rangle = \frac{\langle m_x {\mathcal H} \rangle}{\langle m_x \rangle}- \langle {\mathcal H} \rangle.
\end{align}
\end{subequations}

The specific heat $C$ [Eq.~\eqref{eq.c}] and the derivatives of the order parameters [Eqs.~\eqref{eq.chi}, \eqref{eq.D0}, and \eqref{eq.D1}] are useful for determining transition points and their universality classes. In particular, the extremum of each observable $\mathcal{O}$ as a function of $T$ defines a finite-size estimate of the transition temperature $T_{{\rm max}}^{\mathcal{O}}(L)$ (the so-called $L$-dependent pseudo-transition temperature), which is known to converge into the transition temperature in the thermodynamic limit. In the case of the second-order phase transition, the leading asymptotic behavior is
\begin{align}
  T_{{\rm max}}^{\mathcal{O}}(L) - T_c \propto L^{-1/\nu}.
\end{align}
Also, the extremum of each observable $\mathcal{O}$ at $T = T_{{\rm max}}^{\mathcal{O}}(L)$ is known to scale with $L$ as
\begin{subequations}
\begin{align}
  \label{eq.scal_c}
  C^{}_{{\rm max}}(L) 
  &\sim
  \begin{cases}
    c_0 + c_1 L^{\alpha/\nu}\  & \text{for $\alpha \neq 0$},
    \\
    c_0 + c_1\ln L\ & \text{for $\alpha = 0$}, 
  \end{cases}
  \\[3pt]
  \label{eq.scal_chi}
  \chi^{}_{x,{\rm max}}(L)
  &\propto L^{\gamma/\nu},
  \\[3pt]
  \label{eq.scal_dm}
  d^{}_{1/T}\,m^{}_{x,{\rm max}}(L) 
  &\propto L^{(1-\beta)/\nu},
  \\[3pt]
  \label{eq.scal_dlm}
  d^{}_{1/T}\ln m^{}_{x,{\rm max}}(L) 
  &\propto L^{1/\nu},
\end{align}
\end{subequations}
where $x=$ A, B, and $o$ distinguishes the different observables defined above, which may diverge at different phase transitions.
From the above FSS relations one can estimate the critical exponents $\alpha$, $\beta$, $\gamma$ and $\nu$, and thereby determine the corresponding universality class.

The above FSS arguments are not applicable in a straightforward manner in the case of the TLIA model (which we have, say, for $\JAB = 0$), simply because it shows no LRO down to zero temperature. However, it is known that the ground state displays quasi-long-range ordering (QLRO) with the spin-correlation function showing the power-law decay~\cite{step}:
\begin{equation}
  \label{PL}
  \langle \sigma_{i}\sigma_{j} \rangle 
  \propto 
  e^{i\mathbf{Q}\cdot r_{ij}}
  r_{ij}^{-\eta},
\end{equation}
with $\eta=1/2$
and $\mathbf{Q} = (4\pi/3,0)$.
The power-law decay of the spin-correlation function is also characteristic of the Berezinskii-Kosterlitz-Thouless (BKT) phase~\cite{kost1,kost2}. The exponent $\eta$ can be estimated by FSS of the corresponding order parameter $m_o$, 
which scales as
\begin{equation}
  \label{ms_FSS}
  m_o(L) \propto L^{-\eta/2}.
\end{equation}

\section{%
  \label{sec: GS}
  Ground-state phase diagram
}
\begin{figure}[t]
  \centering
  \if\arXiv0
  \includegraphics[width=0.90\hsize]{FIG03_GS.eps}
  \else
  \includegraphics[width=0.90\hsize,bb=0 0 405 346]{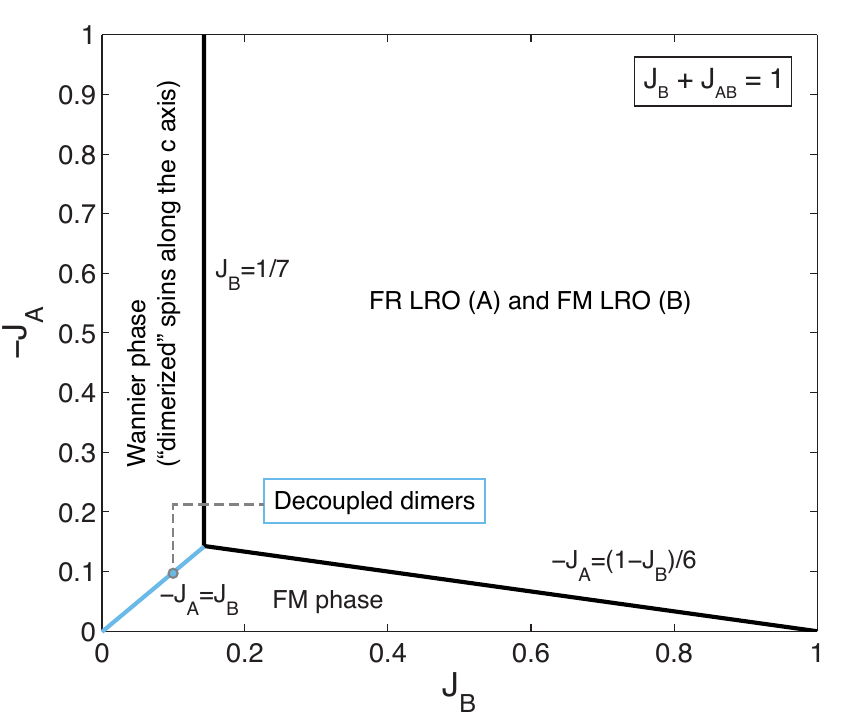}
  \fi
  \caption{%
    \label{fig:GS}
    \if\arXiv0
    (color online) 
    \fi 
    Ground-state phase diagram.
  }
\end{figure}

The ground state (GS) can be determined by considering the energetics of two coupled elementary triangular plaquettes in the adjacent planes; see Fig.~\ref{fig:GS}. We find that for $\JB > (1/6)\JAB$ and $-\JA > (1/6)\JAB$ [$\JB > 1/7$ and $-\JA > (1 - \JB)/6$, respectively, in the unit of $\JB + \JAB = 1$], both planes display LRO: the plane A shows a three-sublattice FR LRO with spins on two sublattices parallel, and those on the third one antiparallel, to the FM ordered spin configuration in the plane B. For $\JB < (1/6)\JAB$, the energetics is dominated by the interlayer coupling $\JAB$, and every NN spin pair coupled by $\JAB$ becomes (anti-)parallel to each other for $\JAB > 0$ ($\JAB < 0$), which we call a ``dimer.'' The in-plane spin configuration is simply determined by the sign of $\JA + \JB$: the GS is the FM state  for $\JA + \JB > 0$ while it has the same massive degeneracy as the TLIA for $\JA + \JB < 0$ (we refer to this as the ``Wannier phase''). Here, the case with $\JA + \JB = 0$ is rather special because the dimers are decoupled at $T = 0$ and we obtain a trivial disordered GS. Finally, the FM GS for $\JA + \JB > 0$ and  $\JB < (1/6)\JAB$ extends to the region where $\JB \ge (1/6)\JAB$, as long as $-\JA < (1/6)\JAB$.

\section{%
  \label{sec:results}
  Finite-temperature results
}

\begin{figure}[t]
  \centering
  \if\arXiv0
  \includegraphics[width=0.90\hsize]{FIG04_PD_v3.eps}
  \else
  \includegraphics[width=0.90\hsize,bb=0 0 405 346]{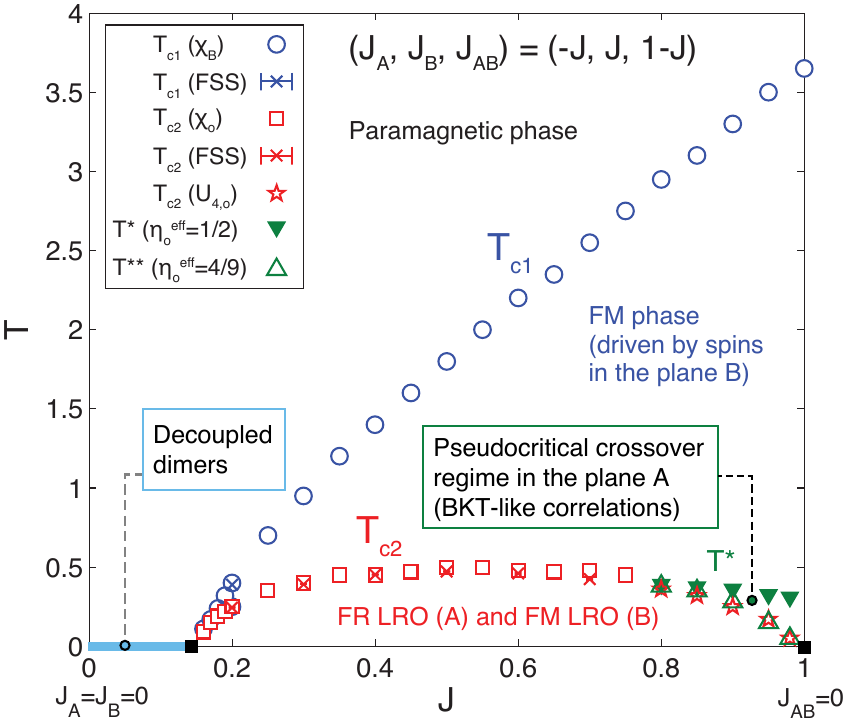}
  \fi
  \caption{%
    \label{fig:PD}
    \if\arXiv0
    (color online) 
    \fi 
    Phase diagram in the $(T,J)$ parameter space
    where $J = -\JA = \JB = 1 - \JAB$. 
    The transition temperatures $T_{c1}$ and $T_{c2}$ are determined in several ways: the maximum of the corresponding susceptibilities (i.e., $\chi_{\rm B}$ for $T_{c1}$ and $\chi_o$ for $T_{c2}$), the crossing of $U_{4,o}$ for $T_{c2}$, and the FSS analysis, as indicated within the parentheses of the legends, with the last one giving most accurate estimates.
    The pseudocritical crossover regime surrounded by the lines of $T^\ast$ at high $T$ and $T^{\ast\ast}$ at low $T$, characterized by $\etaeff_o = 1/2$ and  $\etaeff_o = 4/9$, respectively, is the part of the FM phase in this parameter space ($T^{\ast\ast}$ is expected to coincide with $T_{c2}$ in the thermodynamic limit; see the text).
    The filled squares at $T=0$ indicate the exact interval of the stabilization of the FR-FM ground state, $1/7 < J < 1$. The ground state for $0 \le J < 1/7$ is the trivial disordered state comprising decoupled dimers (see the text).
  }
\end{figure}

\subsection{%
  \label{(sub)sec: -JA = JB}
  Case with $-\JA = \JB$ and $\JAB$ varied
}

\begin{figure*}[t]
  \centering
  \if\arXiv0
  \includegraphics[width=1.00\hsize]{FIG05_-JA-equal-to-JB_observables_v2.eps}
  \else
  \includegraphics[width=1.00\hsize,bb=0 0 1238 685]{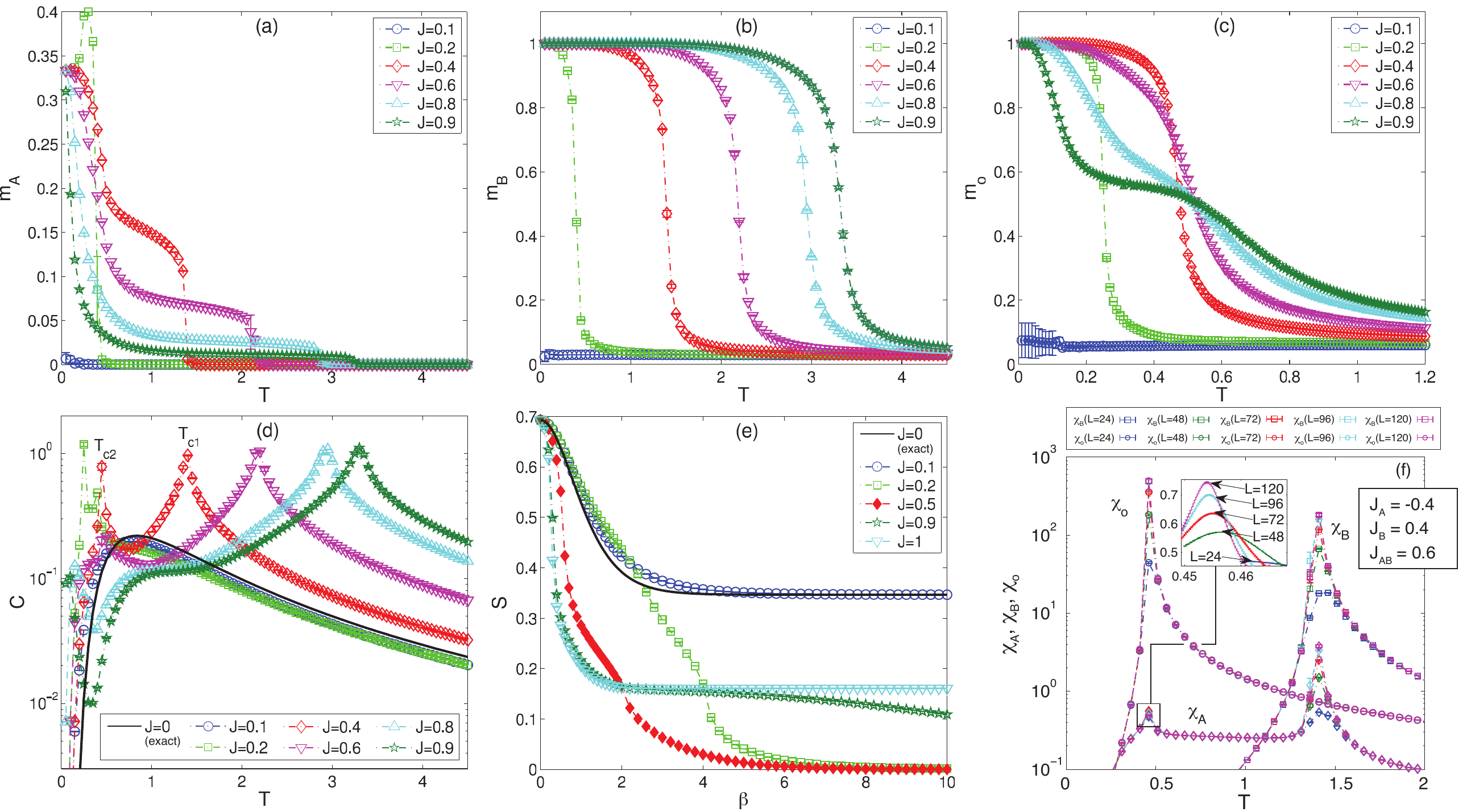}
  \fi
  \caption{%
    \label{fig:-JA-equal-to-JB_observables}
    \if\arXiv0
    (color online) 
    \fi 
    Temperature variations of the sublattice magnetizations 
    (a) $m_{\rm A}$ and (b) $m_{\rm B}$,
    (c) the ferrimagnetic order $m_o$ of the AF layer, 
    (d) the total specific heat, and 
    (e) inverse temperature variations of the entropy density, for various values of $J = -\JA = \JB = 1 - \JAB$ and $L=48$. The solid black curves in (d) and (e) show the exact solutions for the decoupled dimer limit ($J = 0$). $T_{c1}$ and $T_{c2}$ in (d) represent two critical temperatures for $J=0.4$.
    (f) $T$ dependence of $\chi_{\rm A}$, $\chi_{\rm B}$, and $\chi_o$ for $J = 0.4$.
  }
\end{figure*}

First we investigate the case of $-\JA=\JB=1-\JAB \equiv J$ with $J\in [0,1]$. The above GS scenario is corroborated by temperature variations of the magnetizations in the respective planes $m_{\rm A}$ and $m_{\rm B}$, obtained for various values of $J \in \{0.1, 0.2, 0.4, 0.6, 0.8, 0.9\}$ and fixed $L=48$, as shown in Figs.~\ref{fig:-JA-equal-to-JB_observables}(a) and~\ref{fig:-JA-equal-to-JB_observables}(b). Figure~\ref{fig:PD} shows the finite-$T$ phase diagram that we will discuss in the following. For $J=0.1$, or $(\JA, \JB, \JAB) = (-0.1, 0.1, 0.9)$, it is suggested that both $m_{\rm A}$ and $m_{\rm B}$ remain zero at any temperature in the thermodynamic limit, whereas for the other cases with $J > 1/7$, they saturate to the zero-temperature values of 1/3 and 1, respectively. In the latter cases, there is a FM phase transition driven by spins in the plane B first at some critical value $T_{c1}$, as $T$ is lowered from the paramagnetic phase. For a range of temperatures below $T_{c1}$, spins in the plane A also show some degree of FM ordering, which is induced by the FM LRO in the plane B. At a sufficiently low temperature $T_{c2}$ a FR LRO develops in the AF plane, as evidenced from the order parameter $m_o$ [Fig.~\ref{fig:-JA-equal-to-JB_observables}(c)] as well as the the second sharp peak in the specific heat curves [Fig.~\ref{fig:-JA-equal-to-JB_observables}(d)]. This FR order breaks the translational and the three-fold rotational symmetry of the underlying lattice.

\begin{figure}[t]
  \centering
  \if\arXiv0
  \includegraphics[width=0.66\hsize]{FIG06_fss_Tc1.eps}
  \else
  \includegraphics[width=0.66\hsize,bb=0 0 410 689]{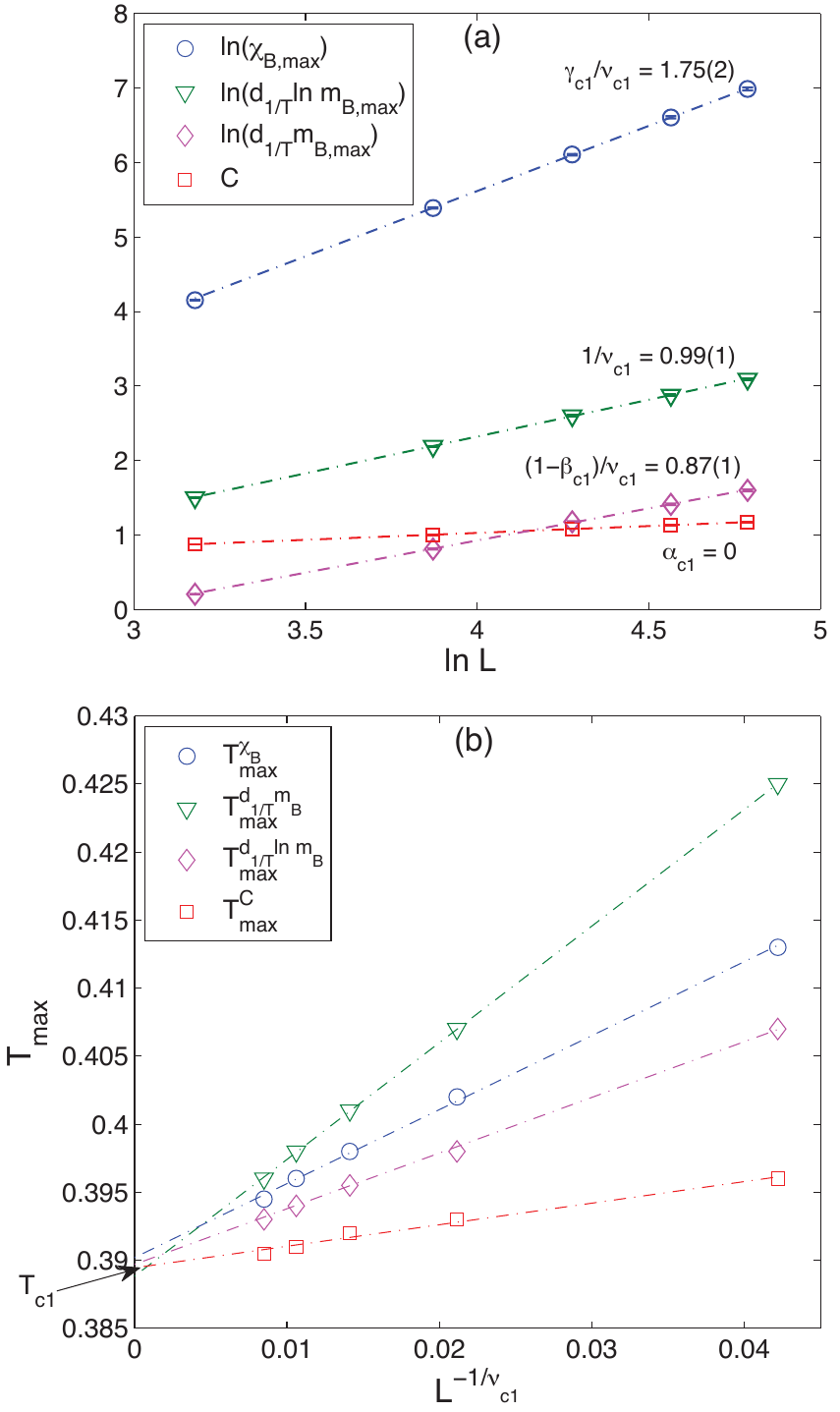}
  \fi
  \caption{%
    \label{fig:fss_Tc1}
    \if\arXiv0
    (color online) 
    \fi 
    (a) Critical exponent ratios and 
    (b) $T_{{\rm max}}^{\mathcal{O}}(L)$
    obtained in the FSS analysis at the high-temperature phase transition into the FM phase driven by spins in the plane B, for $J = -\JA = \JB = 1 - \JAB = 0.2$.
    The arrow in the panel (b) indicates the estimate of $T_{c1}$ in the thermodynamic limit.
  }
\end{figure}

\begin{figure}[h!]
  \centering
  \if\arXiv0
  \includegraphics[width=0.66\hsize]{FIG07_fss_Tc2.eps}
  \else
  \includegraphics[width=0.66\hsize,bb=0 0 410 1374]{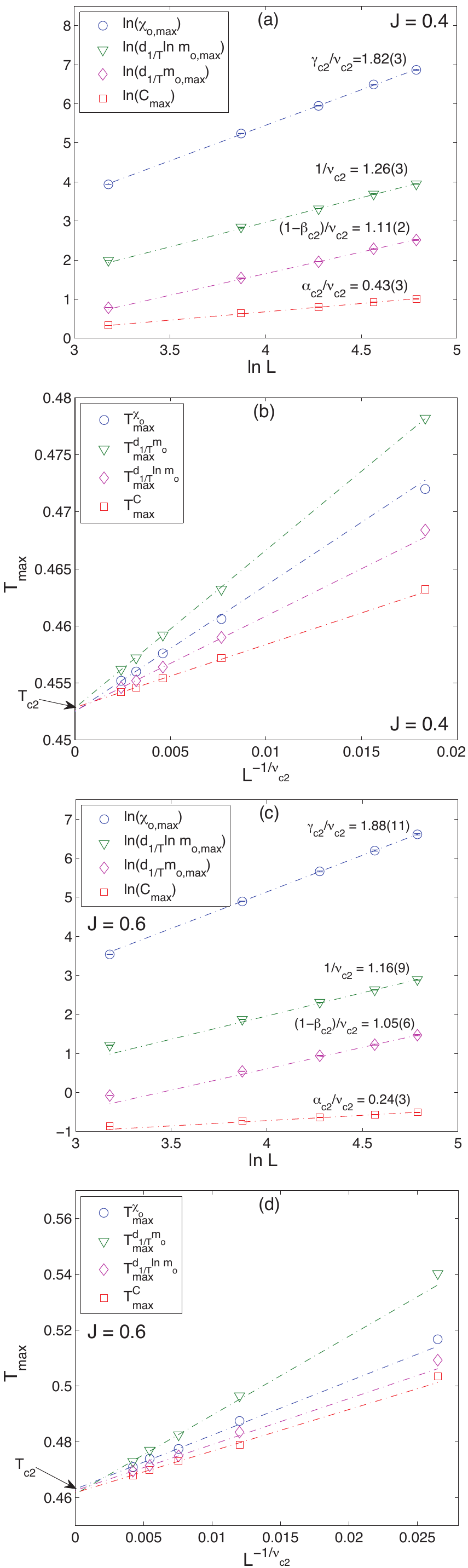}
  \fi
  \caption{%
    \label{fig:fss_Tc2}
    \if\arXiv0
    (color online) 
    \fi 
    (a) Critical exponent ratios and 
    (b) $T_{{\rm max}}^{\mathcal{O}}(L)$ for $J = 0.4$, obtained in the FSS analysis at the low-temperature phase transition into the FR phase ($J = -\JA = \JB = 1 - \JAB$).
    (c) Critical exponent ratios and 
    (d) $T_{{\rm max}}^{\mathcal{O}}(L)$ for $J = 0.6$. The arrows in (b) and (d) indicate the estimates of $T_{c2}$ in the thermodynamic limit for $J = 0.4$ and $J = 0.6$, respectively.
  }
  \vspace{-5pt}
\end{figure}

\begin{figure}[t]
  \centering
  \if\arXiv0
  \includegraphics[width=0.85\hsize]{FIG08_J-dep-exponents.eps}
  \else
  \includegraphics[width=0.85\hsize,bb=0 0 426 693]{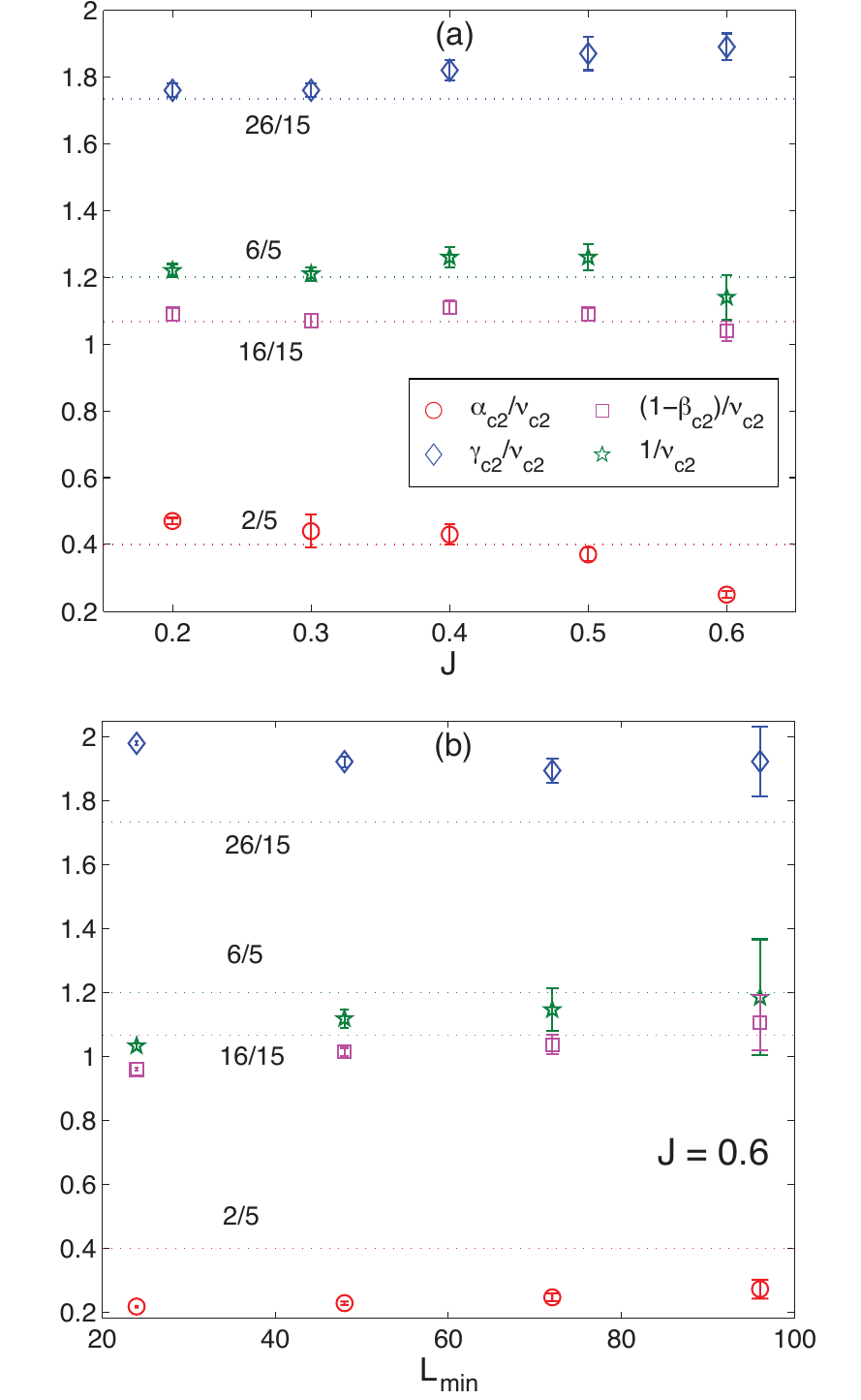}
  \fi
  \caption{%
    \label{fig:indices}
    \if\arXiv0
    (color online) 
    \fi 
    (a) Critical exponent ratios obtained for various $J =-\JA=\JB = 1 - \JAB$ at the FR transition at $T = T_{c2}$, based on FSS using the lattice sizes $L \geq L_{\rm min}=72$. 
    (b) Evolution of the values for $J=0.6$ when the data for $L< L_{\rm min}$ are gradually dropped from FSS. The dashed lines indicate the universal values of the three-state ferromagnetic Potts model.
  }
\end{figure}

\begin{figure}[t]
  \centering
  \if\arXiv0
  \includegraphics[width=0.85\hsize]{FIG09_pseudo-BKT.eps}
  \else
  \includegraphics[width=0.85\hsize,bb=1 2 422 693]{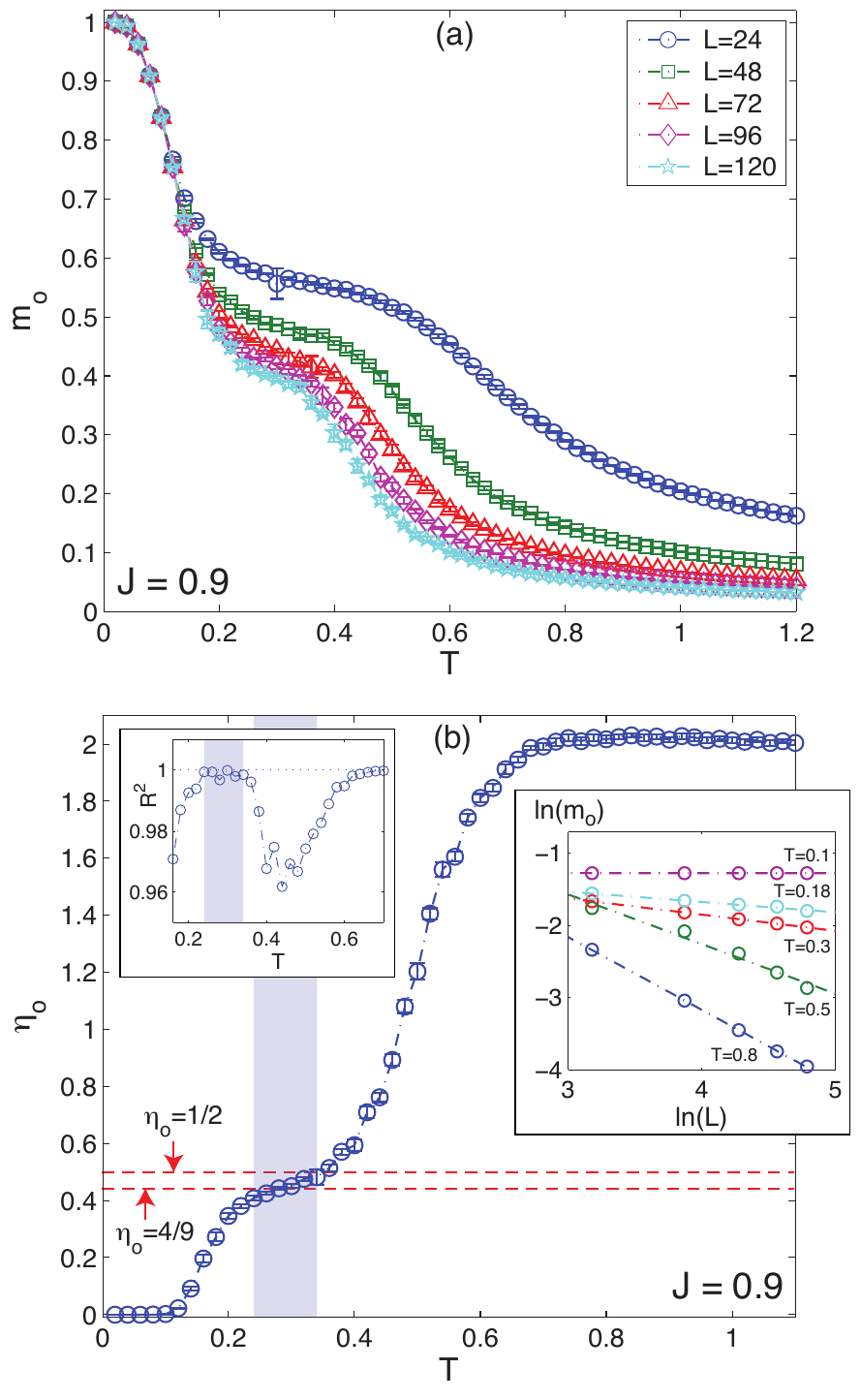}
  \fi
  \caption{%
    \label{fig:pseudo-BKT}
    \if\arXiv0
    (color online) 
    \fi
    Temperature variation of 
    (a) the order parameter $m_o$ for various $L$ and 
    (b) the exponent $\etaeff_{o}$ obtained from the scaling relation Eq. (\ref{ms_FSS}), for $J=-\JA=\JB = 1 - \JAB=0.9$. In (b), the value $4/9 \le \etaeff_{o} \le 1/2$ is expected for the BKT-like spin-correlation in the plane A induced by a coupling to the FM ordered layer B (see the text in Sec.~\ref{sec:crossover}). The insets show the FSS log-log plots of $m_o(L)$ for some representative values of $T$ (right) and the coefficient of determination of the linear fit $R^2$ as a function of $T$ (left); the region with $R^2 \approx 1$ at low temperatures is indicated by the shaded area here and also in the main plot of (b).
  }
\end{figure}

The nature of the respective phase transitions can be studied by performing the FSS analysis. The high-temperature FM phase transition driven by spins in the plane B is identified as a second-order transition belonging to the Ising universality class ($\alpha_{\rm I} = 0$, $\beta_{\rm I} = 1/8$, $\gamma_{\rm I} = 7/4$, $\eta_{\rm I} = 1/4$, and $\nu_{\rm I} = 1$). No significant deviation from the standard behavior is found even in the vicinity of $J = 1/7$, as shown in Fig.~\ref{fig:fss_Tc1}(a) for $J=0.2$. The corresponding critical temperature is obtained from the FSS in Fig.~\ref{fig:fss_Tc1}(b), as $T_{c1}=0.3895(6)$. Thus, it appears that the coupling of the FM layer to the frustrated AF layer results in lowering of the transition temperature but otherwise does not alter the universality class. On the other hand, the universality class of the low-temperature FR transition is clearly different from the behavior of the TLIA model. As shown in Fig.~\ref{fig:-JA-equal-to-JB_observables}(c), the order parameter $m_o$ starts to increase at this transition point, which means that the three-fold symmetry is broken in the FR phase (the preemptive enhancement of $m_o$ seen for $J = 0.8$ and $0.9$ will be discussed shortly). In Fig.~\ref{fig:fss_Tc2}(a) we present the FSS results for the respective critical exponents for $J=0.4$. In fact, the estimated values of $1/\nu^{}_{c2}=1.26(3)$, $\alpha^{}_{c2}/\nu^{}_{c2}=0.43(3)$, $(1-\beta^{}_{c2})/\nu^{}_{c2}=1.11(2)$ and $\gamma^{}_{c2}/\nu^{}_{c2}=1.82(3)$ are quite close to the universal values of the three-state ferromagnetic Potts model~\cite{alex75,kinz81}, with the exact critical exponent ratios given by $1/\nu_P=1.2$, $\alpha_P/\nu_P=0.4$, $(1-\beta_P)/\nu_P=1.0\bar{6}$ and $\gamma_P/\nu_P=1.7\bar{3}$. The critical temperature is estimated in Fig.~\ref{fig:fss_Tc2}(b) as $T_{c2}=0.4527(2)$ for $J = 0.4$. Similar values of the critical exponent ratios are obtained for other values of the parameter $J$, as shown in Fig.~\ref{fig:indices}(a). We find that as $J$ approaches larger values (namely, as the interplane coupling becomes smaller), the exponents appear to deviate from the Potts values; see Fig.~\ref{fig:fss_Tc2}(c) for $J=0.6$, for which $T_{c2} = 0.4622(8)$ [Fig.~\ref{fig:fss_Tc2}(d)]. We believe that this is just a finite-size effect and expect the deviation to diminish at larger system sizes. In fact, such a trend can already be observed in the present data if the data for $L < L_{\rm min}$ are gradually dropped from the FSS analysis, as shown in Fig.~\ref{fig:indices}(b) for $J = 0.6$.

At the FR phase transition, there is an anomaly also in the sublattice magnetization $m_{\rm A}$ and the corresponding susceptibility $\chi_{\rm A}$ diverges, as more clearly seen in the inset of Fig.~\ref{fig:-JA-equal-to-JB_observables}(f) showing the reweighting results of $\chi_{\rm A}$. However, the corresponding critical exponent governing this power-law divergence appears to be related to some secondary scaling operators, as the sublattice magnetization $m_{\rm A}$ is not a proper order parameter for this phase transition.

We find that the system shows peculiar behaviors in the specific heat and the entropy just above the FR phase for $J \approx 1$, namely, when $\JAB$ is very small relative to $\absJA$ and $\JB$. For $J = 0.8$ and 0.9, for instance, the specific heat shows a dip between the two peaks [Fig.~\ref{fig:-JA-equal-to-JB_observables}(d)]. As this implies, there is a plateau region of the entropy as a function of $T$ [Fig.~\ref{fig:-JA-equal-to-JB_observables}(e)]. The entropy value in the plateau is close to (1/2) $\times$ 0.3231 $\approx$ 0.1615, which corresponds to the half of the GS value of the single-layer TLIA~\cite{wann50}. Therefore, the peculiar behavior is ascribed to fluctuations in the plane A. As mentioned earlier, the FR order parameter $m_o$ takes finite values for finite $L$ within the corresponding temperature range, but slowly decays with increasing $L$ [see Fig.~\ref{fig:pseudo-BKT}(a)]. These features resemble the characteristics of a BKT-like phase, although we will argue that this is actually a \textit{pseudocritical} crossover regime induced by a proximity to a Gaussian fixed point (see Sec.~\ref{sec:crossover}), with a rather large but finite correlation length for the FR local order parameter $\phi^{}_{R}$ in Eq.~\eqref{eq:phi}, namely, $\langle \phi^{}_{R} \phi^{}_{R'} \rangle \propto \lvert{R - R'}\rvert^{-\etaeff_{o}} \exp(-\lvert{R - R'}\rvert/\xi)$ with large $\xi \gg 1$. Assuming for the moment the algebraically decaying correlation function [Eq.~(\ref{PL})] in this region, the corresponding effective exponent $\etaeff_{o}$ is estimated from the FSS expression [Eq.~(\ref{ms_FSS})]. As shown in Fig.~\ref{fig:pseudo-BKT}(b), the decay in this temperature range for $(\JA, \JB, \JAB) = (-0.9, 0.9, 0.1)$ ($J = 0.9$) is well approximated by a power law with $\etaeff_{o} \approx 1/2$, accompanying a slow monotonic variation of $\etaeff_o \lesssim 1/2$ as a function of $T$. In this sense, this BKT-like behavior is distinct from both the low-temperature FR phase with $\etaeff_{o} = 0$ and the high-temperature paramagnetic phase with $\etaeff_{o} = 2$. In fact, the quality of regression to the power law behavior is excellent within the corresponding temperature range: the coefficient of determination for regression shows $R^2 \approx 1$ within the plateaulike regime of $\etaeff_o$ ($R^2 \approx 1$ at high temperatures corresponds to a paramagnetic behavior), as shown in the left inset of Fig.~\ref{fig:pseudo-BKT}(b). We will examine this peculiar behavior in more detail in Sec.~\ref{sec:crossover}. According to a renormalization group argument presented there, it is adequate to introduce characteristic temperatures $T^{\ast}$ and $T^{\ast\ast}$ based on the criteria $\etaeff_o = 1/2$ and $\etaeff_o = 4/9$, respectively, which are presented in the phase diagram (Fig.~\ref{fig:PD}). The range $T^{\ast\ast} < T < T^{\ast}$ approximately coincides with the range of the plateau (Fig.~\ref{fig:pseudo-BKT}).

\subsection{%
  \label{(sub)sec: -JA neq JB}
  Cases with $-\JA \ne \JB$
}

\begin{figure*}[p]
  \centering
  \if\arXiv0
  \includegraphics[width=0.92\hsize]{FIG10_JB_04_JAB_06_v2.eps}
  \else
  \includegraphics[width=0.92\hsize,bb=0 0 1252 1676]{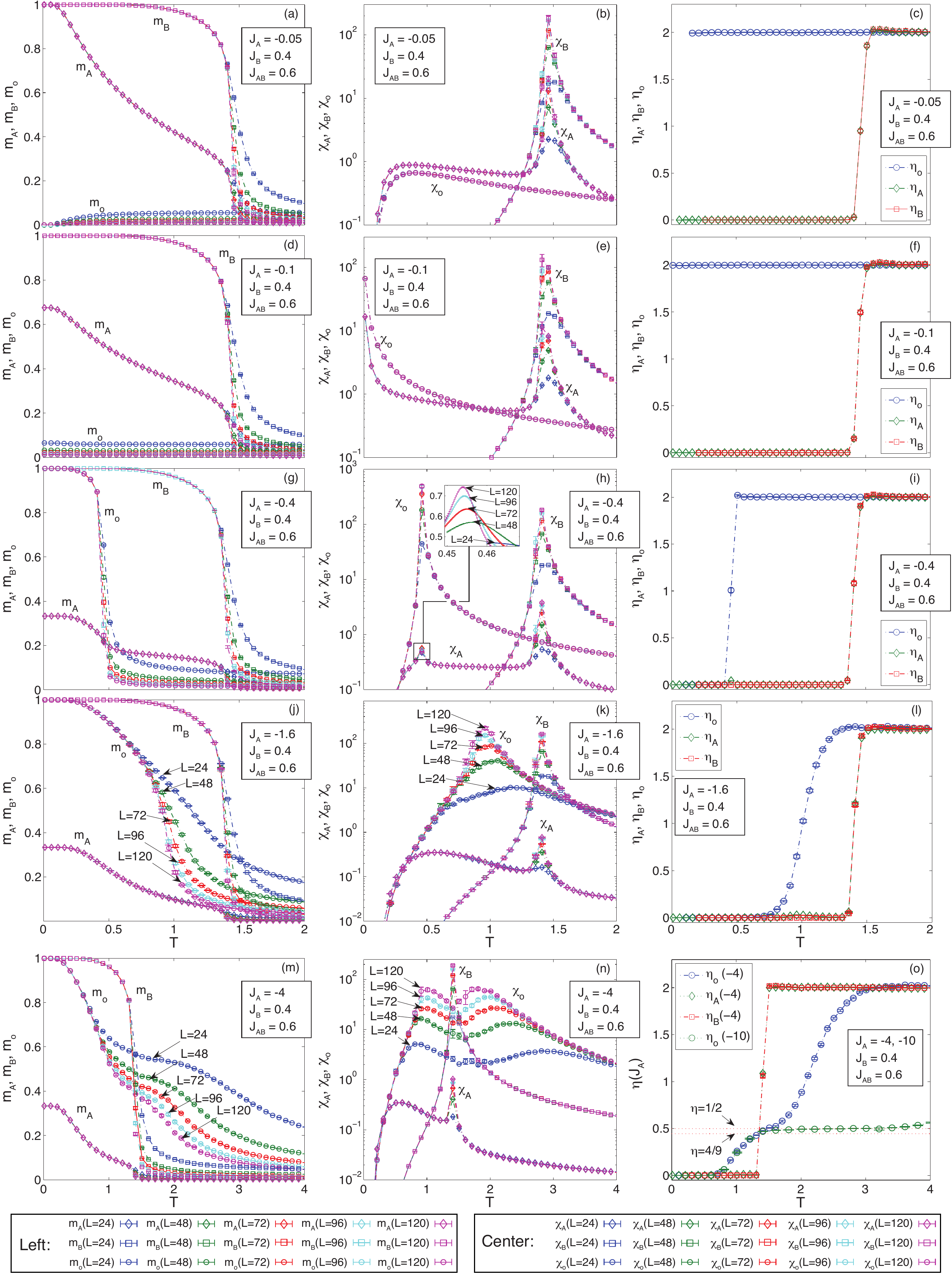}
  \fi
  \caption{%
    \label{fig:JB_04_JAB_06}
    \if\arXiv0
    (color online) 
    \fi Temperature dependencies of 
    the quantities $m_{\rm A}$, $m_{\rm B}$, and $m_o$ (left column), 
    $\chi_{\rm A}$, $\chi_{\rm B}$, and $\chi_o$ (central column), and 
    $\eta_{\rm A}$, $\eta_{\rm B}$, and $\eta_{o}$ (right column) for $\JAB=0.6$, $\JB=0.4$, varying values of $\JA$, and 
    $L=$24--120. The panel (h) is the reproduction of Fig.~\ref{fig:-JA-equal-to-JB_observables}(f) to make a comparison. 
    In the panel (o), $\etaeff_o$ for $\JA = -10$ is also included.
  }
\end{figure*}

\begin{figure}
  \centering
  \if\arXiv0
  \includegraphics[width=0.8\hsize]{FIG11_PD_JB_04_JAB_06_v2.eps}
  \else                                            
  \includegraphics[width=0.8\hsize,bb=0 0 393 348]{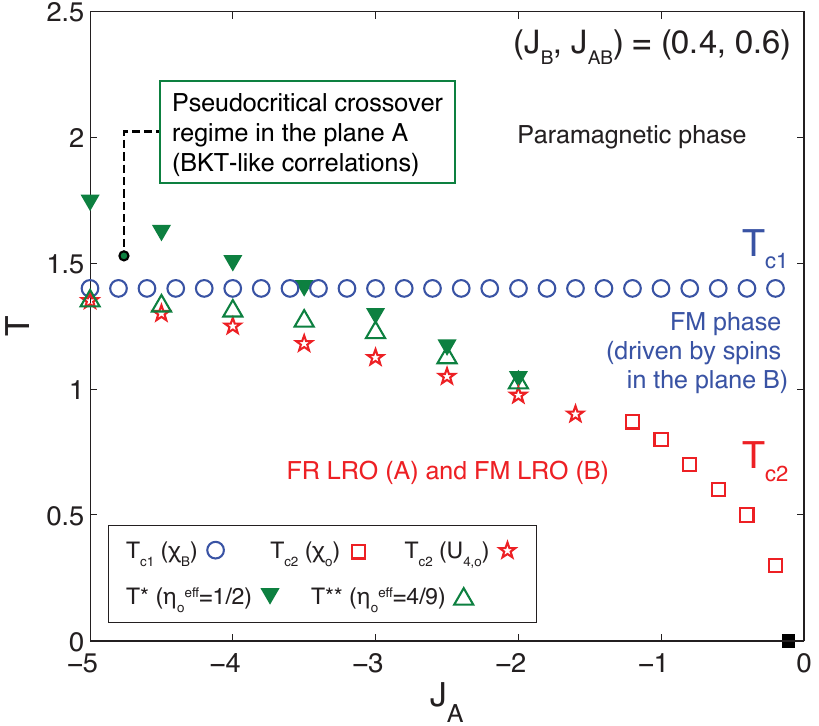}
  \fi
  \caption{%
    \label{fig:PD_JB_04_JAB_06}
    \if\arXiv0
    (color online) 
    \fi 
    Phase diagram as a function of $\JA$ for $(\JB,\JAB) = (0.4, 0.6)$. 
    $T_{c1}$ and $T_{c2}$ are determined by either the maximum of $\chi_{\rm B}$ for $T_{c1}$ ($\chi_o$ for $T_{c2}$) or the crossing of $U_{4,o}$ for $T_{c2}$, as indicated.
    The filled square at $\JA = -(1/6) \JAB = -0.1$ indicates the transition point in the ground-state.
  }
\end{figure}

Next, we discuss more generic cases where $-\JA \ne \JB$. As mentioned earlier, we consider the following two representative cases, $(\JB, \JAB) =$ (0.4, 0.6) and (0.1, 0.9), corresponding to moderate and strong relative strengths of the interplane coupling, and vary $\JA$ in the unit of $\JB+ \JAB$.
For $(\JB, \JAB) =$ (0.1, 0.9), it turns out that the bilayer system can be essentially reduced to the single-layer system of dimerized spins (see Sec.~\ref{sec: GS}), as far as $\absJA$ is of the same order as $\JB$ and not so much exceeding $\JAB$. Within such a range, varying $\JA/\JB$ simply amounts to changing the effective in-plane interaction $\JA + \JB$ for dimerized spins, or block spins in the sense of the Migdal-Kadanoff real-space renormalization group. The corresponding phase diagram is similar to that of the single-layer FM (AFM) model if  $\JA + \JB > 0$ ($\JA + \JB > 0$), and the FM transition in the former case remains in the 2D Ising universality class even in the presence of the coupling to the AF layer.
Thus, below we focus on the case of $(\JB,\JAB) = (0.4,0.6)$. The temperature- and size-dependences of $m_x$, $\chi_x$, and $\eta_x$ with $x=$ A, B, and $o$ are shown in Figs.~\ref{fig:JB_04_JAB_06}(a)--\ref{fig:JB_04_JAB_06}(o) for $\JA \in \{-0.05, -0.1, -0.4, -1.6, -4\}$, in which we include the case of $\absJA = \JB = 0.4$ as a reference to the case considered in Sec.~\ref{(sub)sec: -JA = JB}. Figure~\ref{fig:PD_JB_04_JAB_06} shows the phase diagram that we will discuss in what follows.

We first discuss the cases with $\absJA < \JB$. While we find no significant change in the FM transition at $T = T_{c1}$ compared to the case of $\absJA = \JB = 0.4$, $T_{c2}$ decreases as $\absJA$ decreases and the FR phase vanishes at $\JA = -(1/6)\JAB = -0.1$, where the plane A undergoes a first-order metamagnetic transition accompanying a jump from $m_{\rm A} = 1/3$ to $m_{\rm A} = 1$. For $\JA > -(1/6) \JAB$, the GS of the bilayer system is in the FM phase (Fig.~\ref{fig:PD_JB_04_JAB_06}). The ferromagnetically ordered spins in the plane B induce an effective magnetic field for spins in the plane A, the magnitude of which is $m_B \JAB$ in the mean-field approximation. Thus, reducing $\absJA$ means that this effective field is enhanced relative to the intralayer coupling for the plane A. This explains the observed disappearance of the FR phase.

\begin{figure}
  \centering
  \if\arXiv0
  \includegraphics[width=0.8\hsize]{FIG12_pseudo-BKT-eta_vs_JA.eps}
  \else
  \includegraphics[width=0.8\hsize,bb=0 0 391 467]{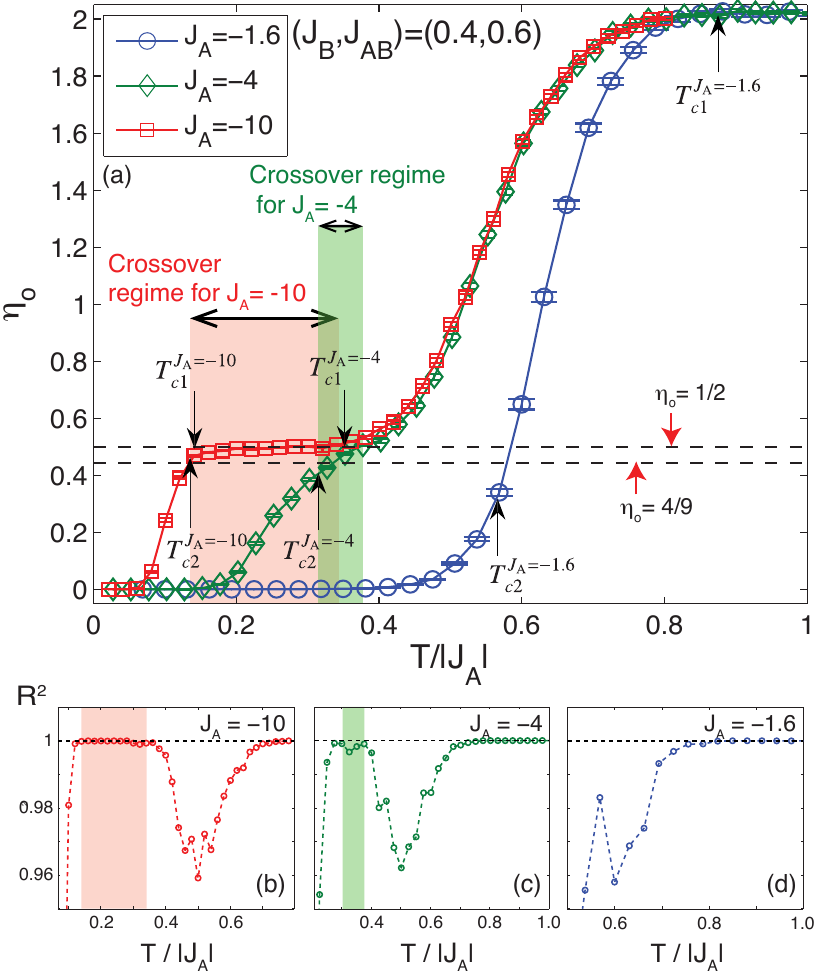}
  \fi
  \caption{%
    \label{fig:nu_mo-T_Jz0_6}
    \if\arXiv0
    (color online) 
    \fi 
    (a) Temperature dependence of $\etaeff_o$ for $\JA = -1.6$, $-4$, and $-10$ with $(\JB, \JAB) = (0.4, 0.6)$.
    (b)--(d) Temperature dependence of the coefficient of determination of the linear fit $R^2$ for each case.
  }
\end{figure}

On the other hand, as we increase $\absJA$ from $\absJA = \JB$, this naturally increases characteristic temperature associated with short-range order (SRO) in the frustrated plane A. For our practical purposes, we can use $T^{\ast}$ (corresponding to $\etaeff_o = 1/2$) also as the temperature scale for this SRO. Figure~\ref{fig:nu_mo-T_Jz0_6}(a) suggests $T^\ast \approx 0.3 \absJA$ for large $\absJA$. In the simulation, the SRO can be seen as enhancement of $m_o$ for small system sizes at $T_{c2} \lesssim T \lesssim T^{\ast}$; see Fig.~\ref{fig:JB_04_JAB_06}(j) for $(\JA, \JA, \JAB) = (-1.6, 0.4, 0.6)$. For sufficiently large $\absJA$, $T^\ast$ can well exceed the FM transition temperature $T_{c1}$, though no LRO can emerge at $T \approx T^\ast$ because of the highly frustrated nature of the interaction. Instead, we find a plateaulike feature of $\etaeff_o \approx 1/2$ for $T \lesssim T^{\ast}$ [see Fig.~\ref{fig:JB_04_JAB_06}(o) where $(\JA, \JB, \JAB) = (-10, 0.4, 0.6)$], meaning that the spin configuration in the plane A in this regime has essentially the same characteristics as the GS configuration of the TLIA~\cite{wann50}, up to a finite but very large length scale that exponentially increases in $\propto \absJA / T$. For smaller values of $\absJA$, e.g., for $(\JA, \JB, \JAB) = (-4, 0.4, 0.6)$, $\etaeff_o$ reveals a similar though much narrower temperature window of $\etaeff_o \approx 1/2$ [Fig.~\ref{fig:JB_04_JAB_06}(o)]. Although we reemphasize that this is a pseudocritical behavior with finite $\xi$, the coefficient of determination for regression shows $R^2 \approx 1$ within the plateaulike regime [Figs.~\ref{fig:nu_mo-T_Jz0_6}(b) and \ref{fig:nu_mo-T_Jz0_6}(c)]. Similar to the case with $\absJA = \JB \gg \JAB$ discussed in Sec.~\ref{(sub)sec: -JA = JB} [e.g., for $(\JA, \JB, \JAB) = (-0.9, 0.9, 0.1)$ shown in Fig.~\ref{fig:pseudo-BKT}(b)], the plateau of $\etaeff_o$ is not completely flat but has a small finite slope. As shown in Fig.~\ref{fig:nu_mo-T_Jz0_6}(a), the finite slope becomes more evident around $T = T_{c1}$. This implies that the small variation of $\etaeff_o$ is induced by a coupling to the FM order in the plane B. In the meantime, it is found that the short-range FR correlation in the plane A is slightly suppressed around the FM transition, but it becomes enhanced again upon further decreasing $T$ below $T_{c1}$. This behavior creates a dip in the $T$-dependence of $\chi_o$ [Fig.~\ref{fig:JB_04_JAB_06}(n)]. 

The system undergoes the FR transition at $T = T_{c2}$. Although this is expected to be in the same universality class as in the case of $\absJA = \JB$, the singularity at the FR transition for $\absJA \gg \JB$ suffers from much more severe finite-size effects; we find, for instance, that the peak of $\chi_o$ at $T \approx T_{c2}$ is more rounded for $(\JA, \JB, \JAB) = (-4, 0.4, 0.6)$ [Fig.~\ref{fig:JB_04_JAB_06}(n)] than in the case of $(\JA, \JB, \JAB) = (-0.4, 0.4, 0.6)$ [Fig.~\ref{fig:JB_04_JAB_06}(h)]. 

When we tune $\absJA$ so that it is still larger than $\JB$ but of more comparable magnitude, the plateaulike feature of $\etaeff_o$ observed for large $\absJA$ increasingly diminishes. Eventually, the plateau disappears and SRO in the plane A develops into the FR LRO without an intervention of a BKT-like temperature window, as in the case of $\JA = -1.6$ [Fig.~\ref{fig:nu_mo-T_Jz0_6}(a)]. In such a case, we find that the FM order in the plane B is established substantially prior to the short-range order in the plane A (namely, $T_{c1} \gg T^{\ast}$). This implies that the FM order parameter $m_{\rm B}$ has to be small enough at $T = T^\ast$ to have an extended region with the pseudocritical BKT-like behavior when $\JB$ and $\JAB$ are of the same order.

\section{%
  \label{sec:crossover}
  Discussion: origin of the pseudocritical BKT-like behavior
}

In the previous section, we investigated two representative cases where the peculiar BKT-like behavior of $\etaeff_o \approx 1/2$ emerges prior to the FR transition in the plane A, accompanying a small monotonic temperature variation of $\etaeff_o \lesssim 1/2$: (i) $\absJA = \JB \gg \JAB$ (Sec.~\ref{(sub)sec: -JA = JB}) and (ii) $\absJA \gg \JB \approx \JAB$ (Sec.~\ref{(sub)sec: -JA neq JB}). In the case (i), the system first undergoes the FM transition at $T = T_{c1}$ driven by spins in the unfrustrated plane B upon cooling, followed by development of SRO in the plane A under the influence of the small interlayer coupling $\JAB$. This SRO is subsequently promoted to the FR LRO at $T = T_{c2}$ and it is during the corresponding ordering process that the BKT-like behavior appears in the plane A. In the case (ii), on the other hand, the large $\absJA$ induces SRO first in the plane A around $T \approx T^{\ast}$ upon cooling. The BKT-like behavior emerges as the FM order is subsequently developed in the plane B, which in the same time gradually affects the spin correlation in the plane A through $\JAB$. In both cases, the spin correlation in the plane A in the BKT-like regime is very well described by a power law (the coefficient of the determination of regression is $R^2 \approx 1$ within the corresponding regime). This extends at least up to the length scale of our largest system size ($L = 120$), or possibly by order of magnitude larger than this in some cases; see below, where we evaluate $\xi / L$ in an effective model up to $L = 768$.

\begin{figure}
  \centering
  \if\arXiv0
  \includegraphics[width=0.70\hsize]{FIG13_in-field-TLAI.eps}
  \else
  \includegraphics[width=0.70\hsize,bb=0 0 408 693]{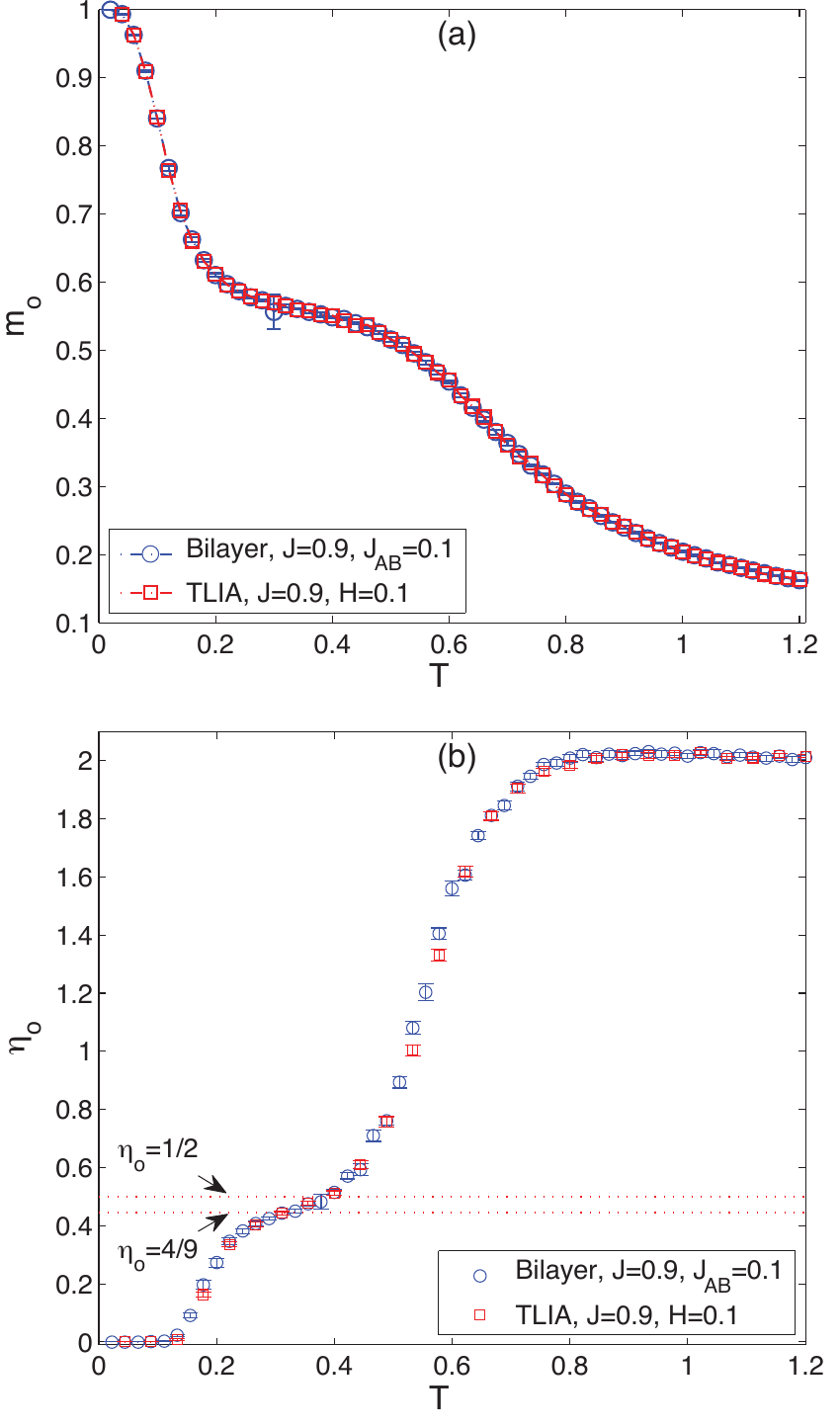}
  \fi
  \caption{%
    \label{fig:TLIA_vs_BL}
    \if\arXiv0
    (color online) 
    \fi 
    (a) The order parameter $m_o$ for $L = 24$ and 
    (b) the effective critical exponent $\etaeff_{o}$, for the bilayer model (blue circles) and the single-layer TLIA model in the field $\tilde{H}=0.1$ (red squares).
  }%
\end{figure}

The emergence of such BKT-like behaviors might be unexpected because the broken symmetry group in the FR GS is $Z_3$, and it is known that the ($p \le 4$)-state clock model in 2D does not support an intermediate critical phase with emergent U(1) symmetry~\cite{Jose1977}. The essential difference from this oversimplified picture is the proximity of the system (more precisely, the plane A) to the degenerate GS manifold of the TLIA, which is under the influence of a small coupling to the FM plane B. As mentioned earlier, the coupling to the plane B can be regarded as an effective magnetic field $\heff \approx m_B \JAB $ in the mean-field approximation for spins in the plane A. This observation motivates us to invoke an effective model description for spins in the plane A by neglecting fluctuations in the plane B~\footnote{%
  For the sake of simplicity, we neglect the possible effect due to short-range FM order in the plane B, which may exist in the case (ii) at $T \ge T_{c1}$. In other words, our discussion applies only to the crossover at $T < T_{c1}$ where $m_{\rm B} \ne 0$.
}%
, namely, by considering a \textit{monolayer} TLIA in a magnetic field defined by
\begin{equation}
  {\cal H}_{\rm TLIA} = -\tilde{J} \sum_{\langle ij\rangle}\sigma_i\sigma_j - \tilde{H} \sum_{i}\sigma_i,
  \label{eq:Heff}
\end{equation}
where $\sigma_i = \pm 1$ denotes an Ising spin at site $i$ of the triangular lattice, representing a spin in the plane A in the bilayer model, and the summation $\langle ij\rangle$ runs over nearest neighbors. The correspondence of the coupling constants is $\tilde{J} \sim \JA$ and $\tilde{H} \sim \heff$. To demonstrate the effectiveness of the mapping, we perform MC simulations of ${\cal H}_{\rm TLIA}$ for $\tilde{J} = -0.9$ and $\tilde{H} = 0.1$ to compare the results with those for the bilayer model with $(\JA, \JB, \JAB) = (-0.9, 0.9, 0.1)$. We focus on the temperature range at $T \ll T_{c1}$ so that we can safely assume $\tilde{H} \sim \heff \approx \JAB$ ($m_{\rm B} \approx 1$). Indeed, as shown in Fig.~\ref{fig:TLIA_vs_BL}, both the order parameter $m_o$ and the critical exponent $\etaeff_{o}$ almost coincide with the respective results for the bilayer model.

The in-field TLIA model~\eqref{eq:Heff} has been investigated rather extensively in the literature~\cite{Kinzel1981,Nienhuis1984,Blote1991,Noh1992,Blote1993,qian04}. Induced by the external field, a three-sublattice FR phase emerges with spins in two sublattices pointing parallel and those in the other antiparallel to the field, which can be directly associated with the FR order in the plane A in the bilayer model. This field-induced transition is in the three-state Potts universality class at finite temperature~\cite{Kinzel1981}, in agreement with our numerical results on the FR transition driven by spins in the plane A. Interestingly, however, Nienhuis and coworkers~\cite{Nienhuis1984} showed that the transition \textit{in the zero-temperature limit} belongs to the BKT universality class, based on the mapping to the 2D Coulomb gas~\cite{Kadanoff1977,Jose1977,Blote1982,Nienhuis1984,Nienhuis1984b}. The crossover from the BKT transition to the three-state Potts universality class induced by thermally excited ``vortices'' (plaquettes of three parallel spins) was also investigated~\cite{qian04}.

\begin{figure}[b]
  \centering
  \if\arXiv0
  \includegraphics[width=\hsize]{FIG14_height.eps}
  \else
  \includegraphics[width=\hsize,bb=0 0 540 153]{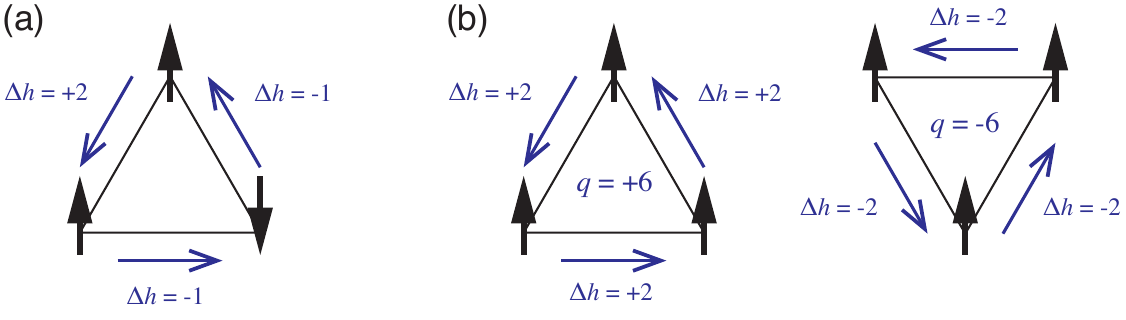}
  \fi
  \caption{%
    \label{fig:height}
    \if\arXiv0
    (color online) 
    \fi 
    (a) The height rule (see the text).
    (b) A plaquette of three parallel spins, which can be three-up or three-down, is a vortex (left) or an antivortex (right) in the height description.
  }
\end{figure}

The observation by Nienhuis \textit{et al.}~\cite{Nienhuis1984} is crucial for explaining the BKT-like phenomena ($\etaeff_o \approx 1/2$ and its small temperature variation with  $\etaeff_o \lesssim 1/2$) observed in the bilayer model. The standard procedure that we follow to describe this physics is to map the TLIA model~\eqref{eq:Heff} onto a height model, also known as the solid-on-solid model~\cite{Blote1982,Nienhuis1984}. By working first on the GS manifold of the zero-field TLIA, namely, by excluding configurations that contain vortices for the moment, we assign an integer-valued height variable $h_i$ to each site $i$ of the triangular lattice. As illustrated in Fig.~\ref{fig:height}, by going counterclockwise around each upward triangle, $h$ changes by $\Delta h = +2$ for parallel spins and by $\Delta h = -1$ for antiparallel spins. This implies that for downward triangles, we should follow the same rule though by going clockwise around them. The fact that the zero-field GS manifold consists of triangles with either up-up-down or down-down-up spins means that the sum over the height increment $\Delta h$ around any single triangle and therefore around any contractible loop is zero. Thus, once the height at the origin is fixed, this prescription leads to a single-valued consistent height map throughout the whole lattice. Here it is convenient to introduce a convention that the height at the origin has to be an arbitrary even (odd) integer if the spin is up (down). Then, the height rule implies that an even (odd) height at any other site also corresponds to a spin up (down), namely, $\sigma_i = \cos(\pi h_i)$. At $\tilde{H} = 0$, this also implies that the height action must be invariant under the global change $h_i \to h_i \pm 1$, $\forall i$. If we take the continuum limit at this point, we obtain the effective action,
\begin{align}
  \mathcal{S}_{\,\text{vortex-free}} = \int d^2 \mathbf{r} \left[ \pi g \left( \nabla h \right)^2 + \sum_{1\le p \le 6} w^{}_p \cos\frac{2\pi h}{p} \right],
  \label{eq:S_v-free}
\end{align}
where $g$ is the stiffness of the height field and the second term contains various periodic potentials; those nonzero in the bare theory are $w_1$ representing the locking potential associated with the discreteness of the height and $w_2 \sim -\tilde{H} / T$ (with $T \to 0$) representing the Zeeman term in the spin model. In addition, the source field to compute $\phi^{}_R$ [Eq.~\eqref{eq:phi}] appears at $p = 6$. $w_{p}$ with $p > 6$ is not allowed because the system is invariant under the global height shift by $\pm 6$~\cite{Blote1982}.

In this notation, the scaling dimensions of the potential terms are $\Delta_p = (2 g p^2)^{-1}$ at the Gaussian fixed point $w_p = 0$. By using $\Delta_{6} = \eta_o / 2 = 1/4$ for the exact solution of the zero-field TLIA, we can calibrate $g = 1/18$ for this case. With this setup, Nienhuis \textit{et al.}~pointed out that the magnetic field term, $w_2$, is irrelevant for small $\tilde{H}/T$ until it becomes marginal at a critical reduced field corresponding to $g = 1/16$. In other words, below a critical value of $\tilde{H}/T$, the effect of the nonzero magnetic field is only to renormalize $g$ in a nonuniversal fashion, leading to a continuous variation of the critical exponent $\eta_o = 2\Delta_6$ within the range,
\begin{align}
  \frac{4}{9} \le \eta_o \le \frac{1}{2}.
  \label{eq:crossover regime}
\end{align}
This is associated with a line of Gaussian fixed points, corresponding to the so-called ``rough'' phase of the height map. At $g = 1/16$, the system undergoes the field-induced BKT transition~\cite{Nienhuis1984,Blote1991,Blote1993,qian04}: $w_2$ is relevant for $g > 1/16$, where the system is in the ``flat'' phase, corresponding to the three-sublattice FR phase. We also note that $w_1$ remains irrelevant within this range.

\begin{figure}[t]
  \centering
  \if\arXiv0
  \includegraphics[width=0.75\hsize]{FIG15_xi_L-T_cmb.eps}
  \else
  \includegraphics[width=0.75\hsize,bb=0 0 407 352]{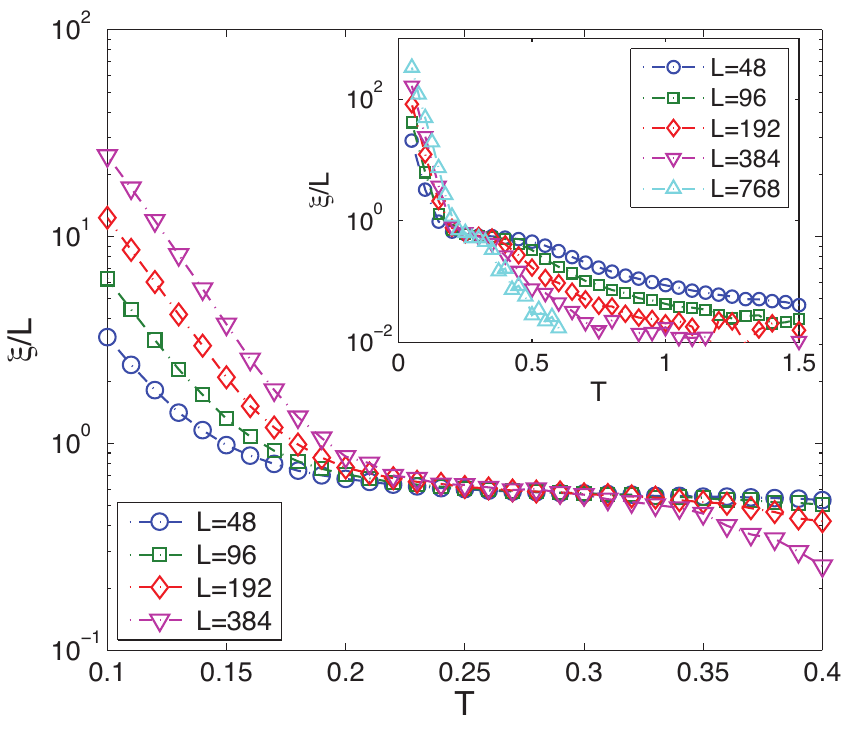}
  \fi
  \caption{%
    \label{fig:xi_L}
    \if\arXiv0
    (color online) 
    \fi 
    Normalized correlation length $\xi/L$, as functions of the temperature, for several system sizes $L$ for $\mathcal{H}_{\rm TLIA}$. The inset shows the behavior in a broader temperature range.
  }%
\end{figure}

\begin{figure}
  \centering
  \if\arXiv0
  \includegraphics[width=\hsize]{FIG16_schematic.eps}
  \else
  \includegraphics[width=\hsize,bb=1 0 535 254]{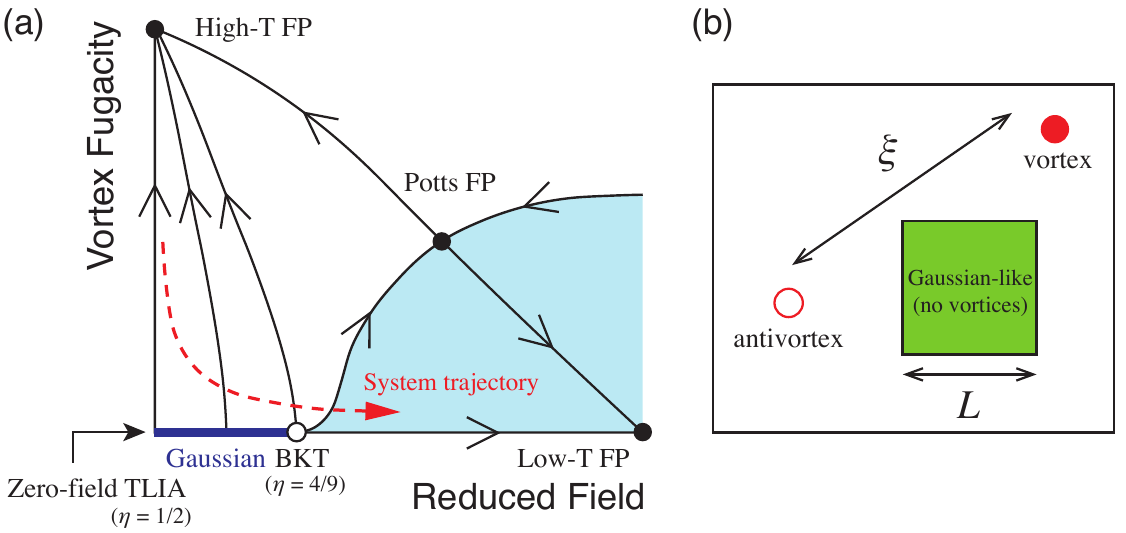}
  \fi
  \caption{%
    \label{fig:schematic}
    \if\arXiv0
    (color online) 
    \fi 
    (a) Schematic projected RG flow diagram including three isolated fixed points (FPs) and a line of Gaussian FPs at $T = 0$. The dashed line indicates a typical trajectory of the system as $T$ is lowered.
    (b) Schematic picture illustrating the situation where the pseudocritical behavior is observed. The system parameters are in proximity of the line of Gaussian FPs and the average vortex-antivortex separation $\approx \xi$ well exceeds the size $L$ of the system (the area represented by the filled square).
  }
\end{figure}

So far, we have restricted our consideration to the GS manifold of the zero-field TLIA. At $T > 0$, vortices and antivortices can be thermally created in pair. As shown in Fig.~\ref{fig:height}(b), they correspond to height dislocations with Burgers vectors $q = \oint dh = \pm 6$, violating the single-valuedness of the height profile. Thus, the meaningful local variable at $T > 0$ is the gradient $\nabla h$ instead of $h$ itself. In the language of the 2D Coulomb gas, these topological defects are magnetic charges whereas the locking potentials discussed above are electric charges~\cite{Kadanoff1977,Jose1977,Nienhuis1984b}. Specifically, the topological defects of $q = \pm 6$ have the scaling dimension $\tilde{\Delta} = 18g$~\cite{Nienhuis1984b}. Hence, within the range $1/18 \le g \le 1/16$, these topological defects remain relevant perturbation to the Gaussian fixed point. In fact, by evaluating the correlation length $\xi(L)$ of the order parameter $m_o$ by MC simulations of $\mathcal{H}_{\rm TLIA}$, we are able to detect both the BKT-like behavior at short distances and the subsequent crossover induced by vortices at larger distances. Here, we evaluate the second-moment correlation length defined by
\begin{equation}
  \label{xi_L}
  \xi(L) = \frac{1}{2\sin(\pi/L)}
  \sqrt{\frac{S(\mathbf{Q})}{S(\mathbf{Q} + \Delta\mathbf{q}_L)} - 1},
\end{equation}
where $S(\mathbf{q})$ is the spin structure factor with $\mathbf{Q}$ and $\mathbf{Q} + \Delta\mathbf{q}_L$ the ordering wave vector and its closest wave vector for the given system size $L$, respectively. As shown in Fig.~\ref{fig:xi_L}, while the dimensionless measure $\xi(L)/L$ for several system sizes falls onto a single line for $0.22 \lesssim T \lesssim 0.3$, as is suggestive of a BKT-like behavior, this region becomes increasing narrower for larger $L$. This observation suggests that while the short distance behavior resembles a power-law with a nontrivial exponent, the genuine long-distance behavior is not. Figure.~\ref{fig:schematic}(a) shows the schematic renormalization group (RG) flow diagram (see Fig.~1 in Ref.~\onlinecite{qian04} for a more precise phase digram of $\mathcal{H}_{\rm TLIA}$). We note that this type of RG flow diagram is rather widely seen among similar systems, aside from important differences in details (see, e.g., Refs.~\onlinecite{lin14} and \onlinecite{Otsuka2011}).

Going back to the bilayer model, we propose the following explanation for the observed pseudocritical behavior. First, the observation of $\etaeff_o \approx 1/2$ can be seen as an indication of temperatures that are low enough relative to the excitation gap of topological defects and also as an indication of the smallness of the reduced effective field $\heff/T \simeq m_{\rm B}\JAB / T$ induced by the coupling to the layer B. As is obvious from this observation, the possible reason for the latter is either the smallness of $\JAB$ [as in the case (i) mentioned in the beginning of this section] or that of $m_{\rm B}$ [as in the case (ii) mentioned in the same place for $T \approx T_{c1}$], or the combination of both. However, since the topological defects give rise to the RG relevant perturbation to the Gaussian theory, the genuine long-distance behavior should deviate from the power law, which is the reason why we refer to it as pseudocritical. Nevertheless, the average separation between vortices and antivortices ($\approx \xi$) grows exponentially at low temperatures, meaning that the ``short-range'' behavior under the strong influence of the Gaussian fixed point can actually extend up to a rather large length scale. As $T$ is lowered further, the gradually enhanced $\heff/T$ is expected to give rise to the nonuniversal renormalization of the effective stiffness and hence the variation of $\etaeff_o \lesssim 1/2$. This crossover seems to be the origin of the slow temperature variation of $\etaeff_o$, which approximately corresponds to the range given in Eq.~\eqref{eq:crossover regime}. 
Because $\xi$ becomes even larger at low $T$, the total average number of vortices within the system size can be very close to zero. Thus, the spin correlation in the plane A is expected be almost perfectly dominated by a Gaussian behavior corresponding to the range of Eq.~\eqref{eq:crossover regime} [see Fig.~\ref{fig:schematic}(b)]. We note that this scenario also provides a natural explanation on why the singularity associated with the FR transition is smeared out when there is a preemptive pseudocritical behavior: the critical behavior is suggested to be dominated by the BKT type with very weak singularities if the system is away from the transition point even by a small degree.

\section{%
  Conclusions
}

We studied magnetic and critical properties of an Ising bilayer system corresponding to a heterostructure of frustrated and unfrustrated triangular lattice layers, with antiferromagnetic (AF) and ferromagnetic (FM) intralayer interactions for the layer A and the layer B, respectively, which are coupled by the interlayer interaction $\JAB$. We showed that the interplay of the ordering tendency in the unfrustrated FM plane and the quasi-degenerate low-energy manifold in the geometrically frustrated AF plane leads to intriguing phenomena, not observed in the separate planes. Our results are summarized in the phase diagrams shown in Fig.~\ref{fig:PD} for $(\JA, \JB, \JAB) = (-J, J, 1 - J)$ with $0 \le J \le 1$ and Fig.~\ref{fig:PD_JB_04_JAB_06} for $(\JB, \JAB) = (0.4, 0.6)$ with $\JA < 0$ varied, where we work on the unit $\JB + \JAB = 1$. In addition, in the limiting cases where the planes are strongly coupled together ($\JAB \gg \absJA, \JB$), the ordering behavior of the entire bilayer is governed by the plane with the dominant intralayer coupling. 

In the first case with $\absJA = \JB$, the bilayer system has the FM order below $T_{c1}$ and the ferrimagnetic (FR) order below $T_{c2}$ ($< T_{c1}$) in the AF layer for $J > 1/7$. The FM transition is in the universality class of the 2D Ising model, whereas the transition into the FR state is in the 2D three-state Potts universality class, both of which are consistent with the broken symmetry groups ($Z_2$ and $Z_3$). The order parameter in the former (latter) case is $m_{\rm B}$ ($m_o$). When the interlayer coupling is small enough (i.e., for $J \approx 1$), the system exhibits pseudocritical Berezinskii-Kosterlitz-Thouless (BKT)-like behaviors prior to the FR transition, and the crossover from the BKT transition appears just above the FR phase. In the second case, where $\JA < 0$ is varied with fixed $(\JB, \JAB) = (0.4, 0.6)$, the FR order in the AF plane is replaced by the FM order in the ground state for $\absJA < (1/6)\JAB$, induced by the interlayer coupling to the FM order in the layer B. On the other hand, for $\absJA \gg \JB$, short-range order is formed in the AF plane even before the FM transition takes place driven by spins in the plane B. This provides another route to an extended pseudocritical regime appearing prior to the FR transition. 

In both cases mentioned above, the BKT-like phenomena can be explained by invoking a mean-field treatment of the interlayer coupling. This approach maps the bilayer system to the monolayer AF triangular-lattice Ising model in an effective magnetic field $\heff \approx m_{\rm B} \JAB$ by neglecting spin fluctuations in the plane B. Then, a two-component Coulomb gas treatment~\cite{Nienhuis1984} suggests that the reduced field $\heff/T$ gives rise to a renormalization of the effective stiffness for the AF layer in a nonuniversal fashion, which leads to to the temperature-dependent small variation of the effective exponent $\etaeff_o$ for the spin-spin correlation function within the range $4/9 \lesssim \etaeff_o \lesssim 1/2$. In this way, a line of Gaussian fixed points controls the ``short-range'' behavior, which extends up to $\xi$ exponentially increasing in $\propto\absJA/T$. Meanwhile, the genuine long-distance behavior beyond $\xi$ is affected by thermally excited topological defects, which induce deviations from the pseudocritical behavior towards the one associated with an ultraviolet fixed point.

In the present study we considered the AF/FM bilayer. A further appealing extension could involve multi-layers formed by stacks of a finite number of the AF and FM planes. It would be interesting to see how the ordering effects from the FM layer propagate through the stack of AF layers, the nature of the critical behavior of which may be additionally controlled by its thickness~\cite{lin14}. 

\begin{acknowledgments}
  This work was supported by the Scientific Grant Agency of Ministry of Education of Slovak Republic (Grant No. 1/0331/15) and the scientific grants of Slovak Research and Development Agency provided under contract No.~APVV-0132-11 and No.~APVV-14-0073. Y.K.~acknowledges the financial support by JSPS Grants-in-Aid for Scientific Research under Grant No.~JP16H02206.
\end{acknowledgments}

\bibliographystyle{apsrev4-1}
\bibliography{ref}

\end{document}